\pdfoutput=1
\documentclass[aps,prl,twocolumn,lengthcheck,
showpacs,letterpaper,footinbib,preprintnumbers,amsmath,amssymb]{revtex4}

\usepackage{graphicx}
\usepackage{rotating}
\usepackage{color}
\usepackage{hyperref}
\usepackage{amssymb}
\usepackage{amsmath}
\usepackage{fancyhdr}
\usepackage{epstopdf}

\newcommand{\rme}{\mathrm{e}}
\newcommand{\rmd}{\mathrm{d}}
\renewcommand{\vec}[1]{\boldsymbol{\mathrm{#1}}}
\newcommand{\eref}[1]{(\ref{#1})}
\newcommand{\Eref}[1]{Eq.~(\ref{#1})}

\newcommand{\xic}{\vec{\xi}}
\newcommand{\ang}{\varphi}
\def\vDoppler{\mathrm{p}}
\newcommand{\F}{\mathcal{F}}
\newcommand{\Fth}{\F_{\rm th}}
\newcommand{\Pspace}{\mathcal{P}}
\newcommand{\nx}{n_\mathrm{x}}
\newcommand{\ny}{n_\mathrm{y}}
\newcommand{\mm}{\mathcal{M}}

\def\affilmrk#1{$^{#1}$}

\def\aeih{1}
\def\uwm{2}

\DeclareGraphicsExtensions{.png}

\begin{document}

\pagestyle{fancy}

\rhead[]{}
\lhead[]{}

\title{Exploiting Large-Scale Correlations to Detect Continuous Gravitational Waves}
\author{Holger~J.~Pletsch\affilmrk{\aeih,}}
\email{Holger.Pletsch@aei.mpg.de} 
\author{Bruce Allen\affilmrk{\aeih,\uwm,}}
\email{Bruce.Allen@aei.mpg.de} 
\affiliation{
 \affilmrk{\aeih}Max-Planck-Institut f\"ur Gravitationsphysik (Albert-Einstein-Institut), Callinstra{\ss}e 38, D-30167 Hannover, Germany\\
 \affilmrk{\uwm}Department of Physics, University of Wisconsin -- Milwaukee, P.O. Box 413, Wisconsin, 53201, USA 
}
    
\begin{abstract}
\noindent
Fully coherent searches (over realistic ranges of parameter space and
year-long observation times) for unknown sources of continuous
gravitational waves are computationally prohibitive. Less expensive
hierarchical searches divide the data into shorter segments which are
analyzed coherently, then detection statistics from different segments
are combined incoherently.  
The novel method presented here solves the long-standing problem of how best
to do the incoherent combination. The optimal solution 
exploits large-scale parameter-space correlations in the coherent detection
statistic.  Application to simulated data shows dramatic
sensitivity improvements compared with previously available (\emph{ad hoc}) 
methods, increasing the spatial volume probed by more than $2$ orders of
magnitude at lower computational cost.
\end{abstract}


\maketitle
 
\textbf{Searching for CW Sources. {\rm---} } 
Direct detection of gravitational waves is the most significant
remaining test of Einstein's General Theory of Relativity, and will
become an important new astronomical tool.

Rapidly rotating neutron stars are expected to generate continuous
gravitational-wave (CW) signals via various
mechanisms 
\cite{owen-1998-58,ushomirsky:2000,cutler:2002-66,jones-2002-331,owen-2005-95}.
Most such stars are electromagnetically invisible, but might be detected 
and studied via gravitational waves.  Recent simulations of neutron star
populations~\cite{knispel-allen-2008, horowitz-kadau-2009} suggest
that CW sources might eventually be detected with new instruments such
as LIGO~\cite{ligo1,ligo2}.  World-wide efforts are
underway to search for CW signals~\cite{pshS4:2008,S4EAH,eahurl} and
observational upper limits already place some constraints on neutron
star physics~\cite{crab-2008,powerflux:2009}. 

Because the expected CW signals are weak, sensitive data analysis
methods are needed to extract these signals from detector noise.  A
powerful method is derived in Ref.~\cite{jks1}. This scheme is based on
the principle of maximum likelihood detection, which leads to coherent
matched filtering.  Rotating neutron stars emit monochromatic CW
signals, apart from a slowly changing intrinsic frequency.  But the
terrestrial detector location Doppler-modulates the amplitude and
phase of the waveform, as the Earth moves relative to the solar
system barycenter (SSB).  The parameters describing the signal's
amplitude variation may be analytically eliminated by maximizing the
coherent matched-filtering statistic~\cite{jks1}.  The remaining
search parameters describing the signal's phase 
are the source's sky location, frequency and
frequency derivatives. The resulting coherent detection statistic
is called the \mbox{$\F$-statistic}.

This work considers isolated CW sources whose frequency varies
linearly with time in the SSB frame.  The corresponding phase
parameter-space~$\Pspace$ is four-dimensional. Standard ``physical''
coordinates on $\Pspace$ are the frequency~$f(t_0)$ at some fiducial
time $t_0$, the frequency's first time derivative~$\dot f$, and a unit
vector $\vec n = (\cos \delta \, \cos \alpha, \cos \delta \, \sin
\alpha, \sin \delta)$ on the two-sphere $S^2$, pointing from the SSB
to the source.  Here $\alpha$ and $\delta$ are right ascension and
declination.  Thus, a point in parameter space
\mbox{$\vDoppler\in\Pspace$} may be labeled by
\mbox{$\vDoppler=\{f(t_0),\dot f, {\vec n}\}$}.  The
\mbox{$\F$-statistic} $\F_\vDoppler[h]$ is a functional of the
detector data set $h$, and is a function of the point in parameter
space \mbox{$\vDoppler\in\Pspace$}.

All-sky searches for unknown CW sources using the
\mbox{$\F$-statistic} are computationally expensive.  For maximum
sensitivity, one must convolve the full data set with signal waveforms
(templates) corresponding to all possible sources.  But the number of
templates required for a fully coherent search increases as a high
power of the observation time.  For one year of data, the computational
cost to search a realistic range of parameter space exceeds the total
computing power on Earth~\cite{bccs1:1998,jks1}.  Thus a
fully coherent search is limited to much shorter observation times.

Searching year-long data sets is accomplished by
less costly hierarchical semicoherent methods \cite{bc2:2000,cutler:2005,hough:2005}.
The data is broken into segments of duration~$T$, where~$T$ is much smaller than one
year.  Each segment is analyzed coherently, computing the
\mbox{$\F$-statistic} on a \emph{coarse} grid of templates.  Then the
$\F$ values from all segments (or statistics derived from $\F$) are
incoherently combined using a common \emph{fine} grid of templates,
discarding phase information between segments.
Among the current semicoherent strategies,  the
Stack-Slide method~\cite{bc2:2000,cutler:2005} sums \mbox{$\F$ values}
along putative signal tracks in the time-frequency plane. 
The Hough transform method~\cite{hough:2005} sums $H(\F-\Fth)$ where $\Fth$ is a
constant predefined threshold. The Heavyside function $H(\mathrm{x})$ is unity
for positive $\mathrm{x}$ and vanishes elsewhere.  This latter technique is
currently used by Einstein@Home~\cite{eahurl}, a public distributed
computing project carrying out the most sensitive all-sky CW searches.

A central long-standing problem in these semicoherent methods is the 
design of, and link between, the coarse and fine grids. 
Current methods, while creative and clever, are
arbitrary and \emph{ad hoc} constructions. This work removes all arbitrariness by
finding the optimal solution for the incoherent step. 
The key quantity is the fractional loss, called \emph{mismatch} $\mm$, 
in expected $\F$-statistic (or sum of $\F$-statistics in the incoherent step) 
for a given signal $\vDoppler$ at a nearby grid point~$\vDoppler'$.  Locally Taylor-expanding
$\mm$ (to quadratic order) in the differences of the coordinates
$\{f(t_0),\dot f, {\vec n}\}$ of $\vDoppler$ and $\vDoppler'$ defines a 
signature ++++ metric $\rmd s^2$ \cite{Sathy1:1996,owen:1996me,bccs1:1998,prix:2007mu}.  
Current methods consider parameter correlations in~$\F$ \emph{to only linear order}
in~$T$.

The $\F$-statistic has large-scale 
correlations~\cite{prixitoh:2005,pletsch1:2008} in the physical
coordinates $\{f(t_0),\dot f, {\vec n}\}$, extending outside a
region in which the mismatch is well-approximated by the metric
%in coordinates $\{f(t_0),\dot f, {\vec n}\}$ 
given above \cite{myfootnote3}.  
Recent work~\cite{pletsch1:2008} has shown that (for a
given signal) the region where the expected $\F$-statistic has maximal
value may be described by a separate equation for each order of~$T$,
when~$T$ is small compared to one year.  The solutions to these
equations are hypersurfaces in $\Pspace$, whose intersections 
are the extrema of (an approximation to) $\F$.

For currently used values of $T$ (a day or longer) it is also crucial to
consider the fractional loss  of $\F$
\emph{to second order} in~$T$~\cite{pletsch1:2008}.  For source frequencies
$\gtrsim 1\,\rm kHz$ and for values of $T \gtrsim 60 \, \rm h$,
additional orders in $T$ would be needed.

\textbf{The new method. {\rm---} }
This work exploits the large-scale correlations in the coherent detection
statistic~$\F$ to construct a significantly improved semicoherent
search technique for CW signals.
The novel method is optimal if the semicoherent detection statistic is taken to 
be the sum of one coarse-grid $\F$-statistic value from each data segment,
and makes four important improvements.

First, the improved understanding of large-scale correlations yields 
\emph{new coordinates} on $\Pspace$. The metric in these coordinates
accurately approximates the mismatch in $\F$ \cite{myfootnote3} in 
each segment. Hence, the optimal (closest) coarse-grid point from each segment 
can be determined for any given fine-grid point in the incoherent combination step.

Second, in the new coordinates we find the first analytical metric for 
the semicoherent detection statistic to construct
the optimal fine grid (minimum possible number of points).
Previous ad hoc approaches obtain the Þne grid by reÞning the coarse 
grid in three dimensions, $\dot f$ and $\vec n$.  
Here, the explicit incoherent-step metric shows that refinement
is only needed in \emph{one} dimension, $\dot f$.
This greatly reduces the computational cost at
\emph{equal detection sensitivity}, although it also reduces the
accuracy of the estimated source parameters.  But
this is a very profitable trade, because in a hierarchical search the
primary goal of the first stages is to discard the uninteresting
regions of parameter space.  Later follow-up stages use longer
coherent integrations to more accurately determine the source parameters.

Third, existing techniques combine the coherent results less
effectively than our method, because they do not use metric
information beyond linear order in~$T$. This leads to a higher
sensitivity at equal computational cost.

Fourth, the new technique can simultaneously do a Stack-Slide-like summing
of $\F$~values and a Hough-like summing of $H(\F-\Fth)$, with a lower
total computational cost than either one of these methods
individually.

For a given CW source with realistic phase parameter
values (\mbox{$f \lesssim 1\,{\rm kHz}$}, \mbox{$|\dot f| \lesssim f/50\,{\rm
    yr}$}) and coherent data segment lengths \mbox{$T\lesssim 60\,\rm h$}, 
the large-scale correlations of the $\F$-statistic are well described
by the first- and second-order (in $T$) equations \cite{pletsch1:2008}: 
\begin{align}
  \nonumber
  &\nu(t) = f(t) + f (t)\dot \xic(t) \cdot {\vec n}  + {\dot f} \,\xic(t) \cdot {\vec n} \,,\\
   \nonumber
  &\dot \nu(t) = {\dot f} + f(t) \ddot \xic(t) \cdot {\vec n} + 2{\dot f} \,\dot \xic(t) \cdot {\vec n}\,,\\
   \label{e:C1}
  &{\rm where}\, f(t) \equiv  f(t_0) + (t - t_0) \dot f\,.
\end{align}
Here $\xic(t)\equiv\vec{r}_{\rm orb}(t)/c$, with 
$\vec{r}_{\rm orb}(t)$ denoting the vector from the Earth's barycenter to the SSB,
and $c$ the speed of light. The
quantities $\nu(t)$ and $\dot \nu(t)$ can be interpreted as the
source's instantaneous frequency and frequency derivative at the
Earth's barycenter at time $t$.

The parameters $\nu$ and ${\dot\nu}$ provide new coordinates on $\Pspace$.
It is also useful to introduce new (real-valued)
sky coordinates $\nx$ and $\ny$ (as in \cite{jk4}):
\begin{equation}
  \label{e:skyparams}
  \nx(t) + i \, \ny(t) = f(t)\, \tau_E \, \cos \delta_D\, \cos \delta \, \rme^{i [\alpha - \alpha_D(t)]} . 
\end{equation}
Here $\tau_E = R_E/c \approx 21\,\textrm{ms}$ is the light travel time
from the Earth's center to the detector, and $\alpha_D(t)$, $\delta_D$
are the detector position at time $t$.  The metric separation $\rmd
s^2$ is
\begin{align}
\label{e:metric}
  \rmd s^2/\pi^2 = & \rmd\nu^2\,  T^2/3 + \gamma^2\, \rmd{\dot \nu}^2\, T^4/180
    + 2\,\rmd \nx^2 + 2\,\rmd \ny^2 \nonumber\\ 
    &- 4\, \rmd \nu\, \rmd \ny\,T/(\pi\ell)    
    + 4\,\rmd \dot{\nu}\, \rmd \nx\,T^2/(\pi\ell)^2\,.
\end{align}
In defining differences in coordinates $\{\nu,\dot\nu,\nx,\ny\}$, the time $t$ in
Eqs.~\eref{e:C1} and \eref{e:skyparams} is the midpoint of the data
segment spanning times $[t-T/2, t+T/2]$, and \mbox{$\gamma = 1$}. To
simplify the form of the metric, $T$ is taken
to be a positive integer number $\ell$ of sidereal days.

The new coordinates $\{\nu,\dot\nu,\nx,\ny\}$ have important
advantages over the original coordinates $\{f,\dot f, {\vec n}\}$.
The metric is explicitly coordinate-independent (showing that
$\Pspace$ is flat). In fact, the region around a point $\vDoppler$ in which the
mismatch $\mm$ is well-approximated by $\rmd s^2$ is \emph{much}
larger \cite{myfootnote3}.

Consider a segment of data $h_{\vDoppler}$ which contains a strong 
CW source with phase parameters $\vDoppler$. 
If the sky separation patch is small
enough to neglect the $\rmd \nx$ and $\rmd \ny$ terms in
\Eref{e:metric}, then $\F_{\vDoppler'}[h_{\vDoppler}]$ is extremized for all
$\vDoppler'$ that have the \emph{same} values of
$\nu$ and $\dot \nu$ as $\vDoppler$. This set of points in
$\Pspace$ forms a two-dimensional surface $\rmd\nu = \rmd {\dot \nu} = 0$.  
Thus, for all sources within the sky patch, there exists a different $(f, \dot
f)$ pair with those same values of $\nu$ and $\dot \nu$.
This property is exploited by the new method.

\textbf{An implementation. {\rm---} }
%To start, 
The data set is divided into $N$ segments of length~$T$
(potentially including short gaps in operation) labeled by the integer~\mbox{$j=1,...,N$}.
The segments span time intervals~\mbox{$[t_j-T/2,t_j+T/2]$}.  The
detector-time midpoint of segment~$j$ is $t_j$ and
$t_0\equiv\frac{1}{N}\sum_{j=1}^{N}t_j$ is the fiducial time.

Every segment is analyzed \emph{coherently} on a coarse grid in~$\Pspace$.  
This grid is constructed so that no
point in $\Pspace$ is farther than a specified distance from some
coarse-grid point, where the distance measure is defined by the metric
of \Eref{e:metric}. To simplify the grid construction, large frequency bands are analyzed 
by breaking them into many narrow sub-bands.
For each segment $j$, and at each coarse grid point, the
$\F$-statistic is evaluated, and ``stats'' are obtained. Here, the
word ``stat'' denotes the two-tuple $(\F_j, H(\F_j-\Fth))$.

For simplicity, the same coarse grid is used for all data segments:
the Cartesian product of a rectangular grid in $f, \dot f$ and a grid on the
sky-sphere ${\vec n} \in S^2$. The spacings are
\mbox{$\Delta f = \sqrt{12m}/(\pi T)$} and \mbox{$\Delta {\dot f} =
  \sqrt{720m}/(\pi T^2)$}, where $m$ is the one-dimensional metric
mismatch parameter \cite{S4EAH}.  The spacing of the coarse sky grid
is chosen so that the $\rmd \nx$ and $\rmd \ny$ terms in
\Eref{e:metric} may be neglected.  When orthogonally projected onto
the equatorial unit disk, the sky grid is uniform, and 
contains $\approx 2 \pi/(\Delta\ang)^2$ points, with $\Delta\ang =
\sqrt{2 m} /(\pi f\,\tau_E\,\cos \delta_D)$.

The \emph{incoherent} step combines the ``stats'' obtained by the
coherent analysis, using a fine grid in~$\Pspace$.  At each point in
the fine grid, a ``stat'' value is obtained by summing one stat value
from each of the $N$ coarse grids.  The coarse grid point is the one
with the same sky position as the fine grid point, which has the
smallest metric separation from \Eref{e:metric}. 
The final result is a ``stat'' value at each point on the fine grid.  The first
element of the stat is the sum of the $\F$-statistic values from the
coarse-grid points.  The second element is a number count, reflecting
the number of data segments in which $\Fth$ was exceeded.  A
detectable CW signal leads to a fine-grid point with a high number
count and a large sum of $\F$-statistics.

The spacing of the fine grid is determined from the metric for the
fractional loss of the expected $\sum_{j=1}^N {\F}_j$ due to parameter 
offsets between a putative signal location and a fine grid point.
%at the fiducial time $t_0$.
This may be calculated as proposed in \cite{bc2:2000}, by averaging the 
coarse-grid metric over the $N$ different segments.
Since each coarse-grid metric is no longer calculated at the 
data-segment midpoints (but at $t_0$), 
the coefficients change between segments because of the 
time-dependence of the parameter-space coordinates.
For our choices of $t_0$ and $T$, the only additional term in the metric
\Eref{e:metric} that does not average to zero is 
\mbox{$(t_j-t_0)^2T^2\rmd{\dot \nu}^2 /3$}.
The averaged metric is the diagonal part of \Eref{e:metric} with
$\gamma^2 = 1+ 60\sum_{j=1}^{N}( t_j-t_0)^2/ (NT^2)$,
where the coordinate offsets 
are calculated at the fiducial time $t_0$.
Thus, the fine grid may be identical to the coarse grid except that
the spacing $\Delta \dot f$ is smaller by a factor~$\gamma$. This is
of order $N$ when the number of data segments is large.  No further
refinement in frequency or sky position is needed.  Coherent
integration over the total observation time would require refining
both $\Delta \dot f$ and $\Delta f$ (increasing points $\propto N^3$),
plus similar sky refinements.

\textbf{Performance comparison. {\rm---} }
Monte Carlo simulations are used to illustrate the improved
performance of this method compared with the conventional Hough
transform technique \cite{hough:2005}. The software tools used are part of
LALApps~\cite{LALApps} and employ accurate barycentering routines with
timing errors below $3\mu$s.  To provide a realistic comparison,
simulated data covered the same time intervals as the input data
used for the current (S5R5) Einstein@Home search \cite{eahurl}. Those
data, from LIGO Hanford (H1, 4km) and LIGO Livingston (L1, 4km),
are not contiguous, but contain gaps when the detectors are not
operating.  The total time interval spanned is about 264 days,
containing 121 data segments of duration $25\,\textrm{h}$ (so approximately $\ell=1$).

The false alarm probabilities are obtained using $5\,000$ simulated
data sets with different realizations of stationary Gaussian white
noise, with one-sided strain spectral density \mbox{$\sqrt{S_h} =
  3.25\times10^{-22}\,{\rm Hz}^{-1/2}$}.  To find the detection
probabilities, different CW signals with fixed strain amplitude~$h_0$
are added.  The parameters \cite{jks1} are randomly drawn from uniform
distributions in cos(inclination~$\iota$), polarization~$\psi$,
initial phase~$\phi_0$, the entire sky, $f(t_0)\in
[100.1,100.3]\,\rm Hz$, and \mbox{${\dot f} \in[-1.29,-0.711]\,\rm
  nHz/s$}.

\begin{figure}[b]
  \includegraphics[height=5.6cm]{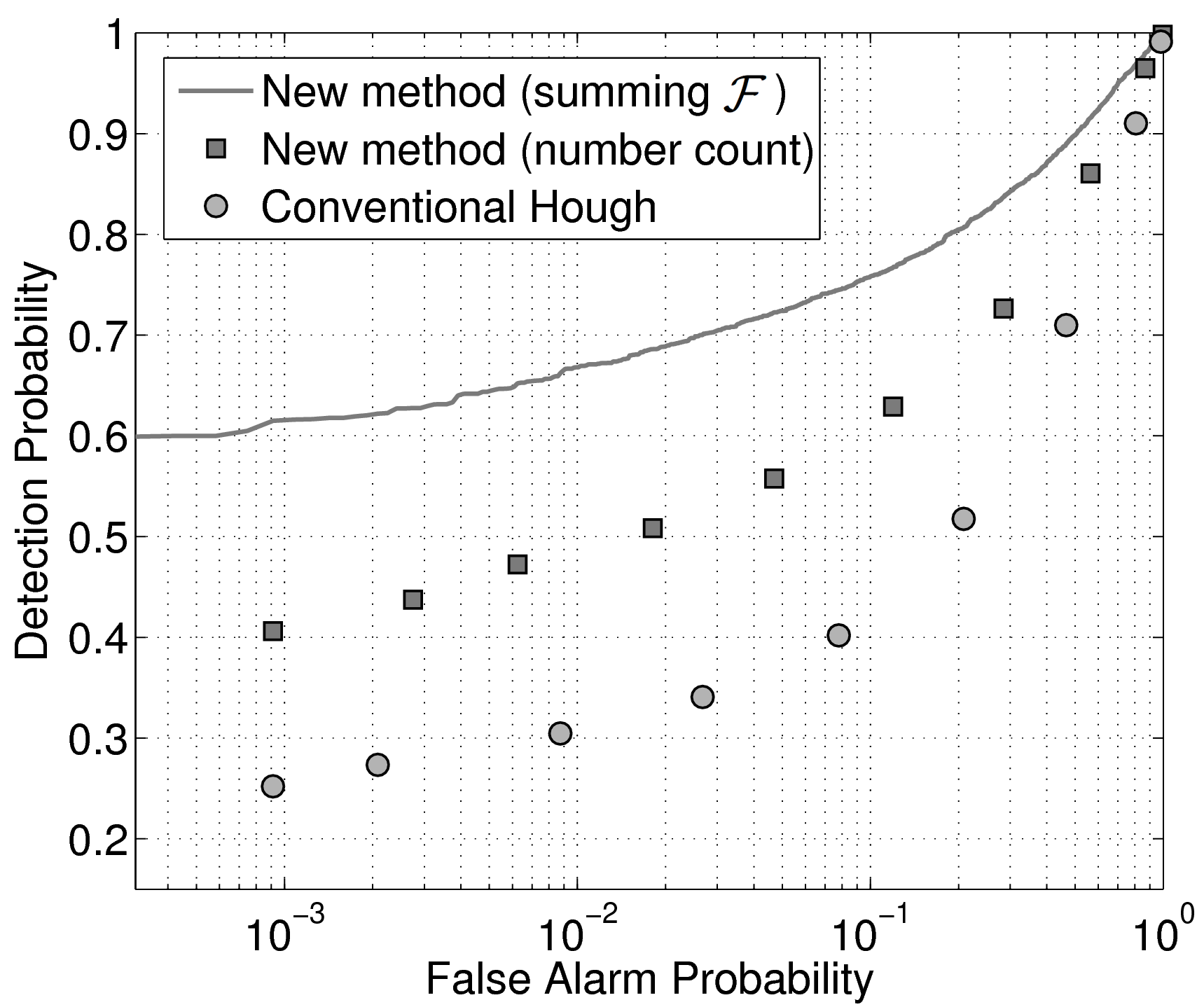}
  \caption{Receiver operating characteristic curve.  The new method performs better than
    the conventional Hough technique.
    \label{f:roc1} }
\end{figure}

Figure~\ref{f:roc1} compares the performance of the different methods.
The receiver operating characteristic is the detection probability
as a function of false alarm probability, at fixed source strain
amplitude~$h_0 = 6\times10^{-24}$.
Because the number count (using $\Fth=2.6$) is discrete, the two
``curves'' in Fig.~\ref{f:roc1} consist of discrete points.  Our method
(using \emph{either} number counts \emph{or} summed $\F$ as a
detection statistic) is superior to the conventional Hough technique.

In addition, this method is computationally faster.  The
comparison used \emph{identical} coherent stages ($m=0.3$, with $2\,981$
coarse-grid points) for both this method and the conventional
Hough.  
But  using \emph{different} fine grids in the incoherent step,
this method's fine grid had $506$ times as many
points as the coarse grid, but the Hough fine grid had $7\,056$ times as
many points.  In spite of using $14$ times fewer fine-grid points,
our method has \emph{substantially higher} sensitivity.

\begin{figure}
          \includegraphics[height=5.6cm]{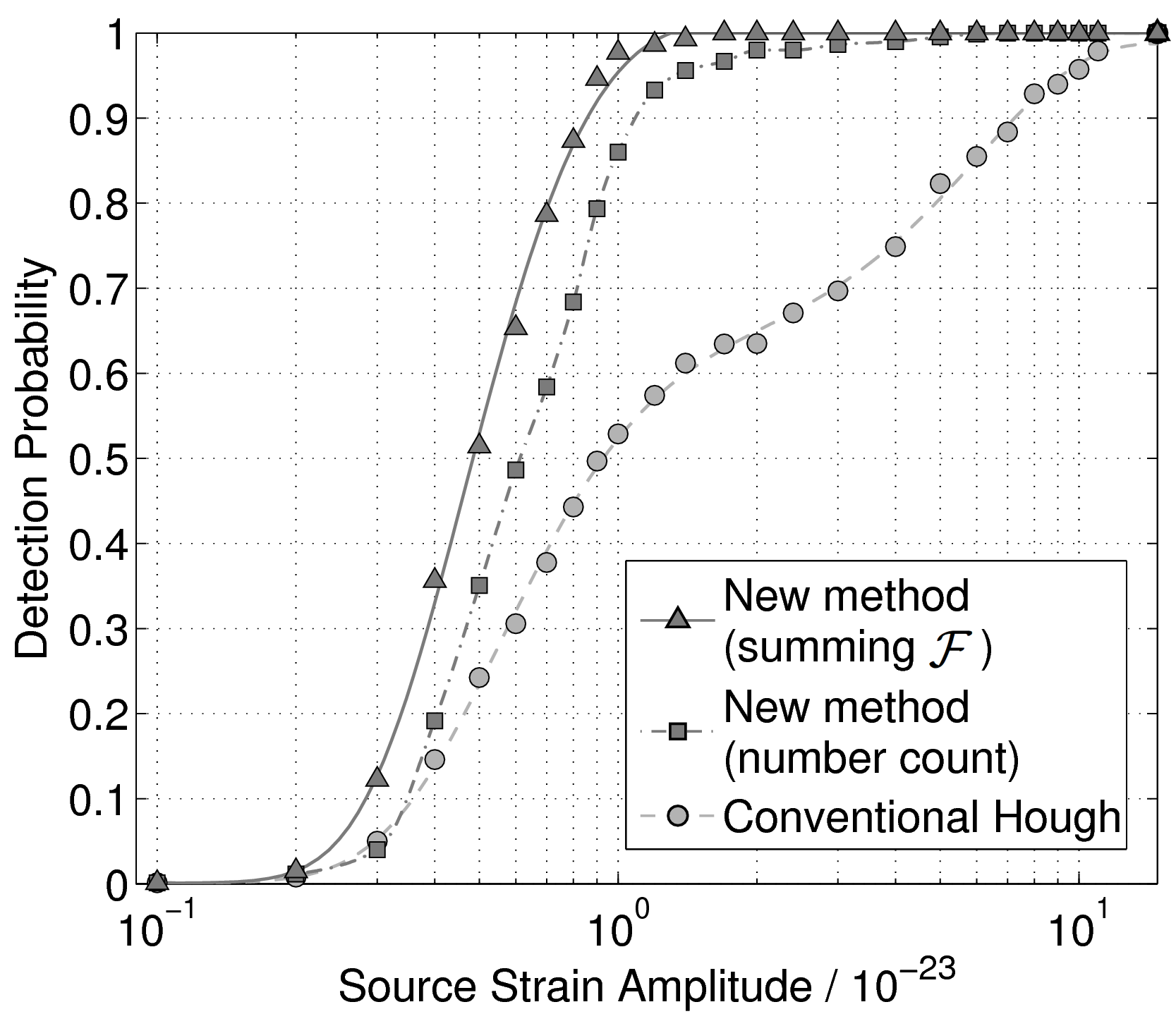}
	 \caption{Detection probability as a 
	 function of source strain amplitude~$h_0$,
	 at $1\%$ false alarm probability. 
	 \label{f:deteff1}}
\end{figure}

Figure~\ref{f:deteff1}  compares the methods.
It shows the detection efficiencies 
for \emph{different} values of source strain amplitude~$h_0$, at a
fixed $1\%$ false alarm probability.  As above, each point in
Fig.~\ref{f:deteff1} is obtained by analyzing $2\,000$ simulated data
sets.  Again, this technique in both modes of operation performs
substantially better than the Hough method.  For example, the
$h_0$ value needed to obtain $90\%$ detection
probability with our technique in number-count
operation mode is smaller by a factor $\approx 6$ than the $h_0$
needed by the Hough method: the ``distance reach''~\cite{jks1}
of our technique is $\approx 6$ times larger.  This increases the
number of potentially detectable sources by more than $2$ orders of
magnitude, since the ``visible'' spatial volume increases as the cube
of the distance.  The lower computational cost of our method
would also allow increases in $N$ or $T$, even further improving the
sensitivity. 

These results are qualitatively independent of frequency, as
confirmed in additional comparisons.

\textbf{Conclusions. {\rm---} } 
A novel semicoherent technique for detecting CW sources has been described.  
In contrast to
previous approaches, the new method exploits large-scale
parameter-space correlations in the coherent detection statistic~$\F$
to optimally solve the subsequent incoherent combination step.
For coherent integration times $T \lesssim 60\, {\rm h}$, the
correlations are well-described by the second-order (in $T$) formulae
presented here.  The method should be extendible to longer
coherent integration times by including higher orders. It could
be extended to search for CW sources in binary systems, 
as well as to space-based detectors.
The method also has applicability in radio, X- and $\gamma$-ray astronomy
(searches for weak radio or $\gamma$-ray pulsars, or pulsations 
from X-ray binaries).

Realistic Monte Carlo simulations show that our technique is much
more sensitive than the conventional Hough method (current
most sensitive all-sky CW search technique).  
The technique presented here is also computationally simpler, and
more efficient.

The LIGO Scientific Collaboration is currently working to deploy this
technique on the Einstein@Home project~\cite{eahurl}, starting with
LIGO S6 data.  The combination of a better search technique,
and more sensitive data, greatly increases the
chance of making the first gravitational wave detection of a CW
source.  The detection of CW signals will provide new means to
discover and locate neutron stars, and will eventually provide unique
insights into the nature of matter at high densities.

We thank R.\ Prix, M.\ A.\ Papa, C.\ Messenger, B.\ Krishnan and B.\ Knispel
for helpful discussions.  This work was supported by the Max Planck
Gesellschaft and by U.S. National Science Foundation Grant No. 0555655 
and No.~0701817.

\bibliography{GCT}

\begin{thebibliography}{27}
\bibitem[{myf()}]{myfootnote3}
\bibinfo{note}{Using $\{\nu,\dot\nu,\nx,\ny\}$ coordinates the metric $\rmd
  s^2$ remains accurate up to $\mm = 0.3$. In contrast, in the $\{f,\dot f,
  {\vec n}\}$ coordinates $\rmd s^2$ can yield errors greater than $10\%$ for
  mismatches as small as $0.001$. In part, this is because the metric in
  $\{f,\dot f, {\vec n}\}$ varies significantly over the $\mm = 0.3$ region.}
\end{thebibliography}

\end{document}